\PassOptionsToPackage{unicode}{hyperref}
\PassOptionsToPackage{hyphens}{url}
\documentclass[
  11pt,
]{article}
\usepackage{xcolor}
\usepackage[margin=1in]{geometry}
\usepackage{amsmath,amssymb}
\usepackage{iftex}
\ifPDFTeX
  \usepackage[T1]{fontenc}
  \usepackage[utf8]{inputenc}
  \usepackage{textcomp} 
\else 
  \usepackage{unicode-math} 
  \defaultfontfeatures{Scale=MatchLowercase}
  \defaultfontfeatures[\rmfamily]{Ligatures=TeX,Scale=1}
\fi
\usepackage{lmodern}
\ifPDFTeX\else
\fi
\IfFileExists{upquote.sty}{\usepackage{upquote}}{}
\IfFileExists{microtype.sty}{
  \usepackage[]{microtype}
  \UseMicrotypeSet[protrusion]{basicmath} 
}{}
\makeatletter
\@ifundefined{KOMAClassName}{
  \IfFileExists{parskip.sty}{%
    \usepackage{parskip}
  }{
    \setlength{\parindent}{0pt}
    \setlength{\parskip}{6pt plus 2pt minus 1pt}}
}{
  \KOMAoptions{parskip=half}}
\makeatother
\usepackage{graphicx}
\usepackage{booktabs}
\usepackage{longtable}
\usepackage{booktabs,tabularx,array,ragged2e}
\usepackage{array}
\usepackage{ragged2e}
\usepackage{authblk}

\renewcommand{\arraystretch}{1.12}
\usepackage{bookmark}
\IfFileExists{xurl.sty}{\usepackage{xurl}}{} 
\hypersetup{
  pdftitle={DigiPhen: a new paradigm for building predictive models of biological systems},
  pdfauthor={H. Steven Wiley; Angela Cintolesi; Niaz Bahar Chowdhury; Jaydeep P. Bardhan; Song Feng; Steven S. Andrews; Herbert M. Sauro; Kristin E. Burnum-Johnson; Scott E. Baker; Douglas Mans},
  hidelinks,
  pdfcreator={LaTeX via pandoc}}

\title{DigiPhen: a new paradigm for building predictive models of
biological systems}
\author[1,2]{H. Steven Wiley\thanks{Corresponding author: steven.wiley@pnnl.gov}}
\author[1]{Angela Cintolesi}
\author[1]{Niaz Bahar Chowdhury}
\author[1]{Jaydeep P. Bardhan}
\author[1]{Song Feng}
\author[2]{Steven S. Andrews}
\author[2]{Herbert M. Sauro}
\author[1]{Kristin E. Burnum-Johnson}
\author[1]{Scott E. Baker}
\author[1]{Douglas Mans}

\affil[1]{Environmental Molecular Sciences Laboratory,
Pacific Northwest National Laboratory,
Richland, Washington, USA}

\affil[2]{Department of Bioengineering,
University of Washington,
Seattle, Washington, USA}

\date{}

\begin{document}
\maketitle

\begin{abstract}

Reengineered biological systems have the potential to revolutionize
chemical and material production, enhance critical mineral recovery,
serve as threat sensors and improve human health. Unfortunately, the
extreme complexity of organisms has made it difficult to achieve this
potential in all but the simplest cases. Recent technological advances,
however, have provided a foundation for solving this problem. Here, we
describe a capability for accelerating the reengineering of cells by
providing accurate predictions of the impact of genetic or environmental
changes on cell phenotype. This ``digital phenome'' platform (DigiPhen)
consists of integrated experimental, analytical and modeling workflows
for building a digital representation of microbial or plant systems. It
is designed around an expanding set of interchangeable, interconnecting
software and experimental modules that can accurately represent the
mechanistic determinants of phenotype. The DigiPhen platform will
systematically collect data on cell composition, spatial organization,
metabolic pathways and regulatory networks in a semi-autonomous fashion
and use this information to build modular, multi-scale models of
biological systems. These models will be used to predict molecular and
environmental changes needed for producing desired biological outcomes.
DigiPhen is intended to be the heart of community research campaigns
that will meet the immediate needs of individual researchers while
fulfilling long-term goals of the scientific community. Altogether, the
DigiPhen platform represents a new paradigm for building predictive
models of biological systems.

\end{abstract}

\section{Introduction}\label{introduction}

Biological systems have the potential to revolutionize chemical and
materials production, isolate rare elements and improve human health
{[}1-3{]}. Because of this potential, an enormous amount of work has
focused on identifying the range of extant organisms and their
functional properties with the aim of either using them directly or
optimizing their molecular machinery for useful purposes. While there
are successful examples of using synthetic biology to optimize metabolic
pathways {[}4{]}, the majority of these cases involve a handful of
well-studied microorganisms. Because of a lack of fundamental biological
knowledge, however, many genetic changes fail to produce an expected
outcome {[}5{]}. One primary reason is the numerous interconnections
between biological networks and the multitude of feedback loops needed
for organismal homeostasis and robustness. Consequently, altering gene
expression or activity might cause changes in unwanted parts of a cell's
regulatory and metabolic machinery as well as alter structural
functionality {[}6{]}.

Assessing the impact of genetic modifications was for decades approached
heuristically, through trial-and-error {[}7{]}. More recently, kinetic
and genome-scale models offered a more systematic approach {[}8{]}.
While these types of models have accelerated biological
discoveries---enabling, for instance, the prediction of essential genes
or metabolic steps that restrain flux under specific conditions {[}9{]}
---the vast complexity of biological systems and their dynamic response
to their environment remains poorly captured. As a result, current
computational models of cellular pathways are mostly descriptive rather
than predictive. Thus, the trial-and-error approach continues to serve
as the only reliable approach for addressing a wide range of biological
interrogations.

The benchmarking of computational models of cellular pathways that is
needed to improve their predictive power, has been hindered by the lack
of standardized data and consistent biological objectives {[}10{]}. This
has created a major bottleneck for progress in the field. Insufficient
information on modification-phenotype relationships, along with limited
modeling frameworks, has further compounded the challenge {[}11{]}.
Solving these problems is hindered by the current social structure of
the biological sciences, where research is predominantly driven by
single investigators or small teams of scientists {[}12{]}. This makes
it difficult to tackle the immense problem of developing broadly
predictive models of cellular behavior. Addressing these longstanding
challenges is well beyond the capacity of any single research group and
therefore requires a new, collaborative, and large-scale approach to
move the field forward. We propose the DigiPhen (Digital Phenome)
platform as one of the foundations of this new approach.

The DigiPhen platform is a community-focused, integrated data
generating-modeling capability whose long-term goal is to provide the
data and modeling foundation for building a microbial ``virtual cell''
{[}13{]}. We have been inspired by the success of the Protein Structure
Initiative (PSI), which laid the foundation for predicting protein
structures from amino acid sequence, a problem that was once thought to
be computationally intractable {[}14{]}. The PSI was initiated in 2000
and focused on developing the technologies needed for high-throughput
protein structure determination {[}15{]}. Although the PSI directly
generated only 5000 3D structures for the Protein Data bank, their
technologies and scientific leadership helped expand this number to
currently over 210,000 {[}16{]}. Significantly, the experimental work
became the foundation for computational predictions of protein structure
based on both evolutionary and thermodynamic rules. This directly led to
the current generation of highly accurate AI systems for predicting
protein structure from its sequence, such as Alphafold2 {[}17{]}. A
large dataset of protein sequence-structure relationships was critical
for the development of AI-powered protein predictions as was the ability
to validate model prediction by experiments. Similarly, an integrated
data generating -- modeling capability is the fundamental paradigm for
DigiPhen with the expectation that it will also lead to systems that can
predict a cell's functional response based on its composition and
environment.

\section{Approach}\label{approach}

The overall strategy in developing the DigiPhen platform was to start
with currently available data generation and modeling technologies that
could be progressively modified to yield an integrated computational and
data-generating framework. The current effort is being led by a
community-focused scientific team at Pacific Northwest National
Laboratory (PNNL) in conjunction with two advisory committees: a
science-focused committee that helps design a yearly research campaign
focused on a specific set of goals, and a data management committee that
is focused on the challenge of ensuring the experiments will generate
the data and metadata needed to support the modeling efforts.

DigiPhen is envisioned as being a major component of the user program at
the Environmental Molecular Sciences Laboratory, one of the premier
Department of Energy Office of Science user facilities of the Biological
and Environmental Research (BER) program
(https://science.osti.gov/ber/Facilities). As such, it must meet the
needs of independent scientific teams performing research on topics
relevant to BER missions. At the same time, DigiPhen will leverage these
research projects to generate the data needed to build large scale,
predictive models of microbial systems.

Predicting cell phenotypes from cell composition is a far more complex
problem than predicting protein structure from its sequence. In
addition, a protein's structure is a more easily defined endpoint than a
cell's ``phenotype''. Previous whole-cell models have been limited to
predicting the impact of a narrow set of conditions and genetic changes
in part because of the lack of data across different environmental
conditions {[}18{]}. The DigiPhen platform is built around a set of
automated, high-throughput instruments that are designed to help fill
this data gap  (\textbf{Table \ref{tab:analytical-capabilities}}), with the 
understanding that current technologies cannot always measure important
cell parameters, such as intermediate metabolites. Continuous
improvement in analytical capabilities is an essential part of the
DigiPhen platform. However, analytical upgrades will be done with an eye
towards the demands of the newest modeling paradigms, helping to drive
progress in both areas.

Despite advancements in methodologies for measuring cell composition
(e.g., proteomics, metabolomics, transcriptomics), these technologies
still do not generate the enormous amount of different types of data
needed to build completely data-driven, genome-scale predictive models
of cell regulatory and metabolic networks {[}13{]}. One way to reduce
the amount of needed data is to constrain solution space through the use
of empirical or physics-based ``rules''. This approach was used
successfully in developing the AI tool AlphaFold2 for predicting protein
structure {[}17{]}. Still, it cost well over \$1B to generate the data
needed to train AlphaFold2 {[}14, 19{]}. Generating the molecular-level
data needed to train a completely data-driven AI model of cell phenotype
would be prohibitively expensive, even if it was possible {[}13{]}.

Creating a higher level of abstraction is a good way to reduce the
complexity of a model and thus the amount of data needed to build it
{[}20{]}. In solving the protein folding problem, the concept of protein
domains was an abstraction used to parse the overall problem into
subproblems {[}21{]}. Evolutionary constraints were also used. Because
protein structure is the target of evolutionary selection rather than
sequence, structures are more highly conserved {[}22{]}. Pairs of
interacting amino acids are also conserved because they contribute to
important structural features {[}23{]}. Thus, there is a definable
relationship between conserved pairs of amino acids and protein
structure. This rule greatly reduced the potential protein folding
landscape and helped make structure predictions computationally
tractable {[}17{]}.

Similarly, there are rules by which biological systems are built and
operate. One generally accepted rule is that biological mechanisms are
both hierarchical and modular {[}24, 25{]}. In complex multicellular
organisms, modularity can be observed in terms of specialized organs and
tissues, such as the heart and lungs {[}26{]}. Even at the cellular
level, a number of highly conserved modules can be identified, such as
mitochondria for energy production, ribosomes for protein synthesis and
the cell cycle for DNA replication and cell partitioning. The
hierarchical regulation of these modules is a yet unsolved problem
{[}27, 28{]}.

At the level of cellular networks, there is even greater ambiguity as to
what constitutes a module {[}29, 30{]}. There are clearly a number of
operational subsystems in cells, such as metabolism, stress responses
and environmental sensing. They are comprised of a series of interacting
protein complexes, but how these are organized, regulated and
coordinated is far less clear {[}29{]}. Recent more detailed studies on
the regulation of metabolic and protein interaction networks have
started to elucidate some of their governing rules, such as the
principle of resource allocation {[}31{]}, which hint at their modular
nature and interconnection. However, far more work is needed to
transform experimental observations into practical rules that can be
used to constrain predictive models {[}29{]}.

Dynamic, multi-omics data generated by the DigiPhen platform will
provide a foundation for building more detailed mechanistic models, such
as kinetic models based on ordinary differential equations (ODEs). These
types of models can be constrained by data on protein abundance and
intercellular metabolite levels {[}32, 33{]}. They can also include
regulatory and feedback mechanisms. Unfortunately, these types of models
tend to be very complex and require detailed knowledge of enzymatic
mechanisms, making them difficult to build and parameterize {[}34{]}.
Creating a level of abstraction, such as modules, is one way to simplify
the building of ODE-based models {[}35{]}.

\begin{table}[!t]
\caption{Current analytical capabilities available to the DigiPhen platform}
\label{tab:analytical-capabilities}
\centering
\small
\setlength{\tabcolsep}{5pt}
\renewcommand{\arraystretch}{1.18}
\begin{tabularx}{\textwidth}{@{}
  >{\RaggedRight\arraybackslash}p{0.18\textwidth}
  >{\RaggedRight\arraybackslash}p{0.34\textwidth}
  >{\RaggedRight\arraybackslash}X@{}}
\toprule
\textbf{Analytical capability} &
\textbf{Instruments} &
\textbf{Generated data} \\
\midrule
Cytometry &
Flow cytometry and fluorescence-activated cell sorting &
Cell numbers, population frequencies, and viability measurements \\
\addlinespace

Plate readers &
Automated optical-density and fluorescence spectrometers &
Cell density, growth rates, reaction kinetics, biomass estimates, cell viability, and enzyme activity \\
\addlinespace

High-content imaging &
Imaging flow cytometry &
Multiparameter phenotype classification, spatial localization, organelle features, and biofilm analysis \\
\addlinespace

Mass spectrometry &
LC--MS/MS (Orbitrap Astral, Orbitrap IQ-X Tribrid, TSQ Altis Plus, and timsTOF) &
Quantitative measurements of intracellular and extracellular metabolites and lipids, protein expression and abundance, protein modifications, and targeted proteomics \\
\addlinespace

Imaging mass spectrometry &
MALDI-MS (timsTOF flex with MALDI-2 and MALDI-12T FT-ICR) &
Protein and metabolite distributions in cells and communities \\
\addlinespace

NMR &
Bruker Avance NMR spectrometer &
Metabolomics, protein structure and conformational analysis, and protein--ligand binding \\
\addlinespace

FT-ICR mass spectrometry &
21-T FT-ICR mass spectrometer &
Ultrahigh-resolution metabolomics, complex lipidomics, natural-product characterization, and modified-protein analysis \\
\addlinespace

Single-cell analysis &
NanoPOTS coupled to high-resolution LC--MS/MS; SPLiT-seq &
Spatial analysis of protein and gene expression \\
\bottomrule
\end{tabularx}

\vspace{3pt}
\begin{minipage}{\textwidth}
\footnotesize
\textit{Abbreviations:} FT-ICR, Fourier-transform ion cyclotron resonance;
LC--MS/MS, liquid chromatography--tandem mass spectrometry;
MALDI-MS, matrix-assisted laser desorption/ionization mass spectrometry;
NanoPOTS, nanodroplet processing in one pot for trace samples;
NMR, nuclear magnetic resonance;
SPLiT-seq, split-pool ligation-based transcriptome sequencing.
\end{minipage}
\end{table}

\section{Foundational Infrastructure}\label{foundational-infrastructure}

Because of the high degree of complexity of biological systems and their
environments, enormous amounts of data will be needed to describe their
responses. Thus the foundation of the Digiphen platform is an automated
experimental workflow coupled to high-throughput measurement
technologies. Data from measurements are automatically linked to
essential metadata and deposited into a central data repository.
Experiments can be set up manually or with agentic AI tools.

A variety of data types will initially be generated from the DigiPhen
platform (\textbf{Table \ref{tab:analytical-capabilities}}). These data
will include optical density (OD) measurements as well as fluorescence
data for microbes with tagged proteins or other fluorescent markers
allowing researchers to generate growth rates for microbes or microbial
communities {[}36, 37{]}. Culture conditions must necessarily be varied
so that a wide parameter space of both nutritional and other (pH,
temperature, etc) growth conditions can be explored. Mass
spectrometry-based multi-omic data will be generated for selected
samples, but choices will have to be made to determine which subset of
cultivations will be used for more detailed phenotyping. Work is in
progress to integrate phenotyping analytics with AI agents to select the
subset of cultivation conditions and/or genetic perturbations to be used
for deeper phenotyping. Standardized data and metadata will be collected
at every step and normalized so that comparisons can be made across
samples and experiments {[}38{]}.

\section{Essential aspects of the DigiPhen
platform}\label{essential-aspects-of-the-digiphen-platform}

\begin{figure}[p]
\centering
\includegraphics[width=\textwidth,height=0.82\textheight,keepaspectratio]{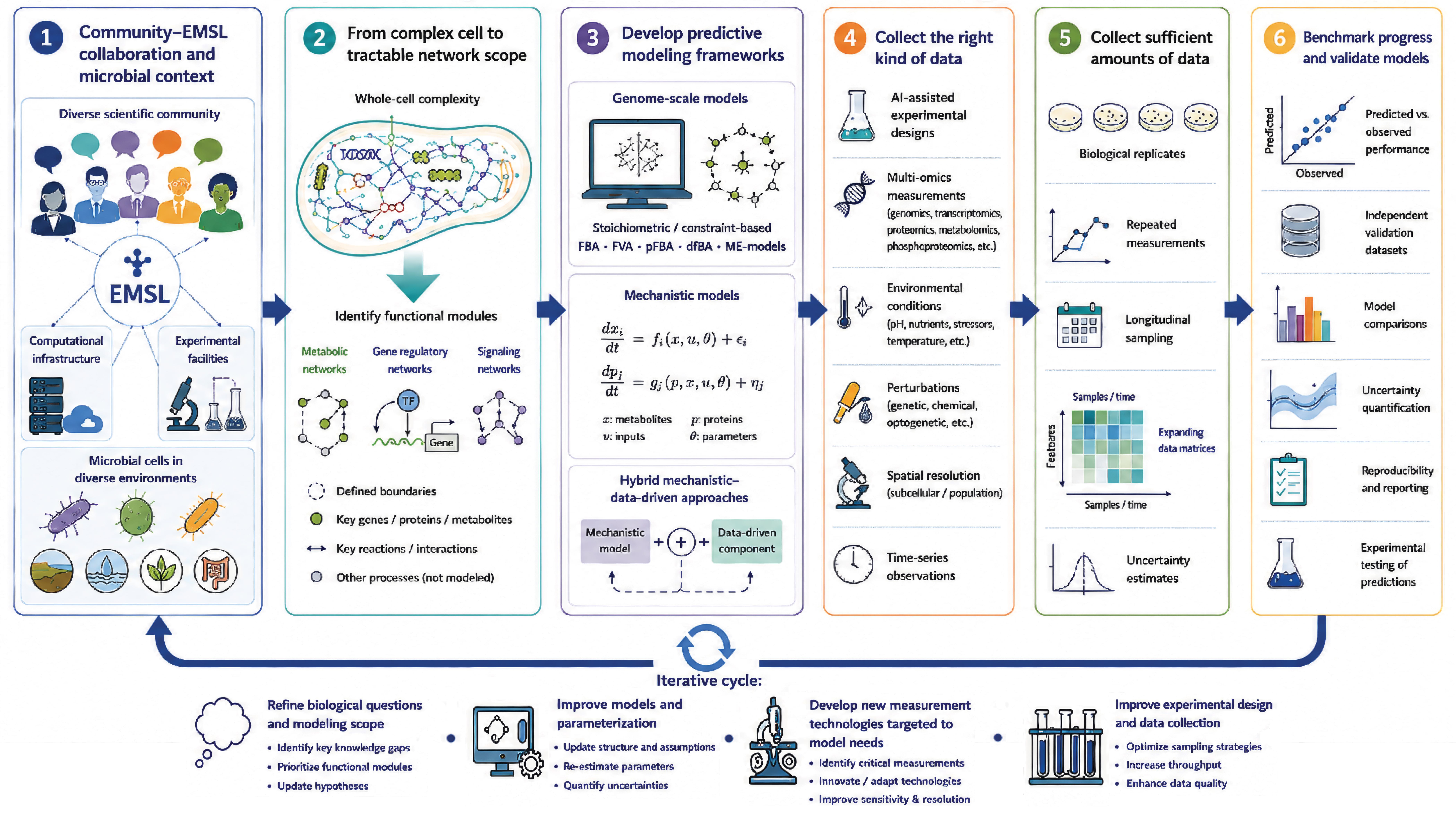}
\caption{A community-driven framework for building a virtual microbial cell. The project is driven by the EMSL user community who brings specific problems that can be addressed by the advanced computational and experimental capabilities in EMSL. Problems typically focus on a specific functional subsystem of cells that can be described in terms of regulatory hierarchy and functional modules. The relationship and activity of these subsystems are described in terms of both genome-scale and mechanistic models. These models are built and parameterized by a wide variety of high-throughput data collected by automated instrument systems. AI-assisted experimental design is used to optimize the data that is collected to enhance model accuracy and usefulness. The resulting models are validated using a community-based benchmarking process. Results of each experimental/modeling campaign are used to iteratively improve the DigiPhen workflow and underlying processes.}
\label{fig:digiphen-framework}
\end{figure}

We acknowledge that developing a useful virtual cell is a daunting task
because of the sheer complexity of biological systems, the lack of
adequate modeling frameworks and a paucity of relevant data. To meet
this challenge, we have designed a multifaceted program, as shown in
Figure \ref{fig:digiphen-framework}. This program is focused on a set of
six distinct goals:

\begin{enumerate}
\def\labelenumi{\arabic{enumi}.}
\item
  Reduce the complexity of the problem
\item
  Develop useful modeling frameworks
\item
  Collect the right kind of data
\item
  Collect sufficient amounts of data
\item
  Engage the scientific community
\item
  Benchmark progress
\end{enumerate}

Other scientific fields have faced similar challenges in building
digital representations of complex systems, and we have designed our
program to incorporate their best practices for dealing with issues such
as complexity and scale. For example, the current large scale earth
system models started from basic weather models that evolved over time
to include ever-increasing spatial and temporal details {[}39{]}. Their
successful evolution over the last 50 years also required community
engagement, data and modeling standards, and rigorous evaluation and
validation of model extensions {[}40{]}. The highly accurate weather
predictions currently generated by these models are a testament to the
success of that approach.

\section{Reducing the complexity of the
problem}\label{reducing-the-complexity-of-the-problem}

Most biological processes are the product of dozens to thousands of
molecular interactions and reactions. Molecular-level models of any but
the simplest systems are not only difficult to build but are
computationally expensive to parameterize and run {[}41{]}. The solution
is to develop a level of abstraction that can hide the complexity
{[}42{]}, such as defining biological networks in terms of their design
patterns or modular structure {[}34, 43{]}.

Modularity is a particularly appealing level of abstraction because it
would directly facilitate building multi-scale models {[}44{]}. It has
also been shown to be a useful concept for reengineering metabolic
pathways {[}45{]}. By definition, the functions of modules are mostly
independent of their context and thus should behave the same regardless
of the system in which they are expressed {[}46{]}. This
context-independence is why modularity is the basis of building
electronic and software system: it allows the behavior of systems to be
predicted from the behavior of their modules rather than individual
components {[}47{]}. This greatly simplifies the process of designing
modified cells for synthetic biology applications.

Although it is widely postulated that biological systems are both
hierarchical and modular, these terms lack precise biochemical
definitions. Several studies have suggested that modules can be defined
in terms of feedback and insulation from downstream elements {[}48{]}.
These concepts have not been widely explored or validated, mostly
because proposed mechanisms of insulation and feedback, such as
molecular modifications and allosteric regulation, are difficult to
experimentally measure. However, recent advances in mass spectrometry
and AI-based tools to infer protein-ligand binding have provided
powerful new approaches to measure biological network structure and
dynamics {[}49-51{]}

An important goal of DigiPhen will be to use new measurement
technologies to help identify boundaries of biological modules and
provide a simplifying level of abstraction for complex biological
systems. Other levels of abstraction, such as functional traits will
also be explored {[}52, 53{]}. The usefulness of such abstractions will
be assessed by the degree to which they can reduce the amount of data
needed to train predictive models of cell responses.

\section{Develop useful modeling
frameworks}\label{develop-useful-modeling-frameworks}

A cell is more than just its metabolism. Building a virtual cell that
can predict responses to a wide variety of environmental and genetic
change will require more than just modeling metabolism. It will require
modeling environmental detection, regulatory mechanisms, replication
machinery, stress responses and anticipatory mechanisms.

Recasting individual subsystems as modules will be helpful, but there
are also issues of describing how they change across different spatial
and temporal scales. Towards this end, several frameworks have been
recently introduced that are capable of linking multiple cell
subsystems, or multiple biological processes into integrated models
{[}54-57{]} . An alternative conceptual framework has been proposed that
uses agent-based models (ABMs) and natural language statements (cell
rules) to create mathematical models of multicellular dynamics {[}58{]}.
Currently, models built using any of these frameworks lack the
mechanistic details needed to make them more generally useful, but they
are important proof-of concept examples for how to create multiscale
models of biological systems.

All modeling frameworks depend on the availability of data that can
build, constrain and parameterize them. Genome-scale metabolic models
(GEMs) are currently the most useful tool for predicting metabolic
activities in a given cell type {[}59, 60{]}. They are based on an
organism's genome and the potential functions of metabolic genes but
generally lack any explicit regulation {[}61{]}. Thus, they are mostly
useful for predicting the potential metabolism of a cell. There have
been several efforts to modify GEMs to include constraints, based on the
idea that the potential for metabolism is constrained by the cell
resources allocated to producing metabolic enzymes {[}62{]}. These
approaches require data on protein abundance and enzyme activity (\emph{kcat}). The
first can be measured or estimated {[}63{]}, but the latter can only be
inferred. More recent developments have included data on enzyme turnover
and deep learning-predicted enzyme kinetics {[}64{]}. Continuing
advancements in analytical technologies should provide data for
improving flux-balance and other modeling frameworks by introducing
additional constraints and increasing the phenotypic space that can be
explored.

Parsing biological systems into modules also provides an opportunity to
use surrogate models, such as Universal Differential Equations (UDEs)
for subsystems that are poorly understood at a mechanistic level
{[}65{]}. For example, we have used transfer functions generated from
experimental data to successfully represent protein modules within
signaling networks {[}66{]}. Likewise, we have used machine learning
(ML) to predict the relationship between mRNA abundance, codon usage and
translational efficiency even without full knowledge of the underlying
mechanisms {[}67{]}. The large amount of detailed data that will be
generated by the DigiPhen platform will provide a foundation for using
AI/ML to build hybrid models that combine mechanistic and surrogate
components.

Other modeling approaches for building virtual cells include the use of
visible neural networks (VNNs) that use ontologies to assign proteins to
functional groups or nodes rather than allowing random mathematical
links. When trained on the output of several million gene perturbation
experiments, this class of models can accurately map the phenotypic
impact of knockout of any given functional group (e.g., {[}68{]}).
However, nodes in these models lack mechanistic details and only work
within the context of the phenotype-specific trained model. Thus, they
cannot be exchanged with other research groups, limiting their
usefulness to the community at large.

It is essential that models supported by the DigiPhen platform be
designed around community standards. This will allow them to be
incorporated into multiscale models being developed by the wider
research community {[}69, 70{]}. It will also facilitate their use
across different software capabilities. If such standards are not
available, the DigiPhen team will work with the modeling community to
establish them. These standards must be open and ideally based on
underlying biological mechanisms, such as the case with the systems
biology markup language (SBML) {[}71{]}. Being able to assess the
reliability and reproducibility of DigiPhen models will also depend on
the use of community exchange standards, such as SBML {[}70{]}.

\section{Collect the right kind of
data}\label{collect-the-right-kind-of-data}

All models that seek to reproduce biological processes require
sufficient amounts and types of quantitative data for both their
construction and parameterization {[}72{]}. Unfortunately, the ideal
types and amounts of data are rarely available. For example, modeling
enzymatic mechanisms ideally requires data on the in-situ concentrations
and localization of enzymes, substrates and product flux {[}73, 74{]}.
This information is extremely difficult to directly measure, especially
in the quantities needed to parameterize models. A more typical approach
is to collect large amounts of inexpensive data, such as gene transcript
abundance, and then try to infer the properties of interest. This not
only introduces inevitable uncertainties into the models but also places
a large burden on the data analysis process {[}75{]}.

The obvious solution to the lack of appropriate data is to develop
technologies to rapidly and inexpensively measure the desired
information. This is not always possible, at least in the short term,
but it should always remain the goal. In the meantime, improved
methodologies for inferring key parameters will be essential. However,
this requires understanding what types of data are most critical for
building different types of models. For example, simple flux balance
models that only require genomic information are unreliable {[}76{]}.
Adding additional constraints, such as protein resource allocation, can
improve their predictive power {[}77{]}, but requires substantially more
data.

Although there is a substantial amount of biological data already
available in public repositories, much of it is not easily reusable
{[}78{]}. For example, modeling metabolic feedback requires knowledge
from multiple biological levels, including~transcriptional control of
enzyme expression, post-translational modifications, and allosteric
regulation~by metabolites. However, much of the data in public databases
on allosteric regulators is erroneous because of lack of appropriate
assay conditions {[}79{]}. Fortunately, multiple groups have described
new approaches for both measuring enzyme-metabolite interactions and
inferring them based on protein structure predictions {[}80{]}, but the
reliability of these predictions has not been widely validated. Directly
measuring the impact of potential regulators is more reliable {[}79{]},
but much more costly.

To maximize the impact of the DigiPhen platform, there must be a dialog
between modelers, data generators and data analysts to ensure that the
most informative, cost-effective data for building virtual cells being
collected. Identifying the most informative data will be one of the main
research problems that the DigiPhen team will tackle. This will require
an understanding of the best approaches for building predictive models,
the most feasible technologies for generating data for those models and
determining which types of analysis can provide the most reliable
results. It will also be an iterative process where models enabled by
new data types can be used to identify data types that will improve
their performance. Machine learning approaches can also be productively
used to infer important parameters for mechanistic models when they
cannot be directly measured {[}81{]}. Because technology development
will always lag the identification of specific data needed in modeling,
identifying the necessary technologies will be a high priority.

\section{Collect sufficient amounts of
data}\label{collect-sufficient-amounts-of-data}

Both traditional modeling as well as AI-enabled approaches require
substantial amounts of data for their building and training {[}82{]}. In
general, the more complex the model, the larger the amount of data
needed in its construction. For example, mass spectrometry-based
metabolomics involves trade-offs between analytical breadth,
quantitative accuracy, and throughput. Targeted assays can deliver
absolute quantification of a predefined panel of metabolites, whereas
untargeted profiling detects thousands of features but yields largely
relative measurements, of which only a small fraction can be confidently
annotated {[}83{]}. Advances in fast chromatography, ion mobility and
data-independent acquisition have nonetheless made broad metabolite
profiling feasible in cohorts of thousands of samples {[}84, 85{]}.
Sample processing also requires tradeoffs with increases in sensitivity
and specificity of an analysis requiring a longer, more complex sample
processing pipeline {[}86, 87{]}. Thus, generating the enormous
quantities of high-quality, relevant data is likely to be one of the
most significant bottlenecks in building a virtual cell.

To address the data quantity problem, the DigiPhen platform is focusing
on highly automated, parallelized workflows that include AI-assisted
experimental design, experiment execution, sample processing, automated
metadata capture and rigorous QA/QC mechanisms. An emphasis will be
placed on keeping sample size to a minimum to reduce costs. Because of
the anticipated scale of the DigiPhen platform, AI-enabled experimental
workflows will be a core capability {[}88{]}

Several additional approaches will be used to increase data volume,
including a focus on particularly data-rich analyses, such as cell
imaging. In combination with advanced image analysis algorithms, cell
imaging can generate a large amount of information from each sample
{[}89{]}. Because mass spectrometry is also capable of generating a
large amount of data from each sample, a high priority will be placed on
increasing sample throughput for this technology. However, an increase
in raw data generation will also need to be accompanied by a
corresponding increase in automated data capture, processing and
analysis {[}90{]}.

\section{Engage the scientific
community}\label{engage-the-scientific-community}

The goal of building a virtual cell is a very ambitious and complex
problem and thus will require the engagement and participation of a
large community of scientists. This is perhaps the greatest challenge
for DigiPhen. Traditionally, biologists have focused on a small part of
an important problem, because of both the complexity of biological
systems and because funding agencies tend to prioritize short term
problems with well-defined outcomes {[}91{]}. Very few individual
scientists would also have the expertise and range of skills needed to
construct a virtual cell. Because DigiPhen is intended to support both
individual users as well as long range scientific goals, it will be
necessary to construct a scientific program that balances the two needs.

The approach that is being used for DigiPhen is to organize work around Research
Campaigns that are designed to systematically address a subset of the
issues needed to build a digital cell. Research Campaigns will also
address current problems that community scientists are funded to solve.
DigiPhen will provide the data needed by these scientists while
exploiting their expertise and their model systems. EMSL scientists will then
provide the missing data and expertise needed to drive progress in
solving a more general scientific goal. Thus, the needs of the user
research teams will be addressed within the context of an overarching
scientific problem.

An example of an overarching problem is the rational management of
microbial resource allocation {[}31{]}. Because of their small size,
microbes face inevitable tradeoffs in allocating their resources to
produce a necessary phenotype (e.g., product production, proliferation,
mobility, etc.). Evolution drives resource allocation to be optimized
for the particular ecological niche for which microbes are adapted. In
nature, a particular niche will vary over time and microbes must be able
to respond to any new conditions. Significantly, the most abundant
microbes in a natural community are those able to grow on the widest
range of substrates {[}92, 93{]} . Thus, significant internal resources
of evolutionarily successful microbes are allocated to molecular
machinery that is only potentially needed to respond to a changing
environment (i.e., anticipatory regulation: {[}94{]}).

When reengineering microbes to produce a new product, a major challenge
is accommodating the associated metabolic impact. Resources for
anticipatory regulation could ideally be reallocated for this. In
addition, one would like to be able to accurately predict functionally
unnecessary pathways that generate by-products that must be removed
during downstream processing {[}95{]}. This would not only relieve the
metabolic burden of these non-essential pathways, but also greatly
facilitate downstream processing which is frequently the main cost of
biomanufacturing {[}96{]}.

A more efficient approach would be to identify the modules and cellular
subsystems that the cell dedicates to anticipating environmental changes
that cannot occur in a bioreactor (e.g., heat shock) or pathway modules
that use substrates that are absent {[}97{]}. Once identified, such
unnecessary subsystems could be deleted, freeing up metabolic resources
for product production. However, this requires understanding the
modularity of the subsystems involved in these ``unnecessary''
functions. Being able to identify genes that encode currently
unnecessary functional modules could have immediate impact for
reengineering microbes.

Many scientists working on developing microbial strains for
bioproduction are focused on defining optimum growth conditions,
substrate utilization capabilities and stress responses that can
compromise cell growth {[}98{]}. The DigiPhen platform will be designed
to rapidly explore relevant growth conditions using autonomous
experimental cycles, producing data that is most likely to be useful to
a user. However, the DigiPhen platform can also collect data for a much
wide variety of experimental conditions. This might not be of immediate
use to an individual user, but when combined with data from other
evaluated strains would be extremely useful in identifying the general
mechanisms by which cells adapt to a shift in conditions. This is one of
the ways that research campaigns can be used to meet the needs of
individual research teams while providing the foundation for building a
virtual cell.

\section{Benchmark progress}\label{benchmark-progress}

An important component of previous large-scale efforts to build
predictive models was establishing formal benchmarking systems to assess
progress. Examples include the Critical Assessment of Structure
Prediction (CASP) experiment, a biennial competition for tracking
protein structure predictions {[}99, 100{]} and the Coupled Model
Intercomparison Project (CMIP) for evaluating earth systems models
{[}101{]}. These types of model validation efforts require identifying
metrics for how well a model recapitulates observed reality and what
types of predictions are most informative. Here we propose the creation
of CAMP -- Critical Assessment of Microbial Phenotypes -- as the
community challenge in periodically assessing modeling performance and
accuracy.

We are currently a long way from building a virtual microbial cell that
could facilitate the reengineering of microbes for useful purposes, so
benchmarking efforts should initially be incremental in nature. The two
aspects that we will initially focus on is predicting the impact of
modifying gene expression and the impact on changing environmental
conditions. Benchmarking model predictions of genetic changes should
include the impact of deleting or amplifying single and multiple genes
from different functional categories (e.g., metabolism, gene regulation)
as well as introducing entirely new genes. As progress is made on
identifying functional modules, benchmarking can be expanded to predict
the impact of removing or introducing modules as a function of cell
environment. Environmental conditions to be tested can include substrate
addition or deletion as well as stress. As the predictive power of the
models improve, validation efforts are expected to include more complex
perturbations.

Because all of the models and data generated by the DigiPhen platform
will be freely available and based on community standards, they should
be compatible with a wide variety of modeling and data analysis
platforms. This will facilitate the participation of multiple research
groups in the effort to build virtual cells. Thus, benchmarking could
include a competition between predictions of models built by different
research groups, similar to the way the DREAM Challenges are designed
{[}11{]}. For example, the competition could be to identify either
genetic or environmental perturbations that would give rise to a
specific phenotype. These predictions could then be experimentally
tested using the DigiPhen platform and the models ranked on how well
they predict the observed phenotypes.

Other benchmarking approaches could be applied. Regardless, the results
of all of these efforts will be used for model improvement and in
designing future research campaigns. It will also help identify the most
informative types of data, which in turn will drive the development of
new analytical technologies. By including rigorous testing into the
design of the DigiPhen platform, we will help ensure that we are making
the most efficient use of our experimental and modeling resources.

\section{Conclusions and the Future}\label{conclusions-and-the-future}

Building a virtual cell that can be used as the basis for cell
reengineering is a very ambitious project, but one is that becoming
increasingly feasible. It is also ideally suited for National
Laboratories. Like the field of physics a hundred years ago, biology has
been mostly in the discovery phase where the parts of the system are
being identified and the rules by which they work discovered.
Universities were, and still are, the place where most discoveries are
being made, driven by individual research teams focused on a small part
of the overall problem. Taking physics to the next level, however,
required a much larger scientific enterprise that could coordinate the
enormous resources needed to integrate the discoveries and implement the
next steps. National Laboratories were created for this purpose and
their success in revolutionizing physics is undisputed {[}102{]}.

As biology transitions from a primarily descriptive to a predictive
science, a similar effort to marshal large-scale scientific and
technical resources is needed to take biology to the next level.
National Laboratories have been laying the foundation for this work over
the last several decades by building foundational analytical and
computational capabilities. Now is the time when building a virtual cell
is feasible and DigiPhen is one of the first step in the process.

Although DigiPhen is based on current state-of-the-art capabilities, it
is likely that both analytical and modeling technologies will continue
to advance at a rapid pace. Thus, continuous analytical upgrades will be
planned with an eye towards the demands of the newest modeling
paradigms, and vice versa, helping to drive progress in both areas. The
ultimate success of the capability, however, will depend on the
engagement and contributions of the scientific community in the project.

\section{Acknowledgements}\label{acknowledgements}

The work described herein was performed in the Environmental Molecular
Sciences Laboratory, Pacific Northwest National Laboratory, a national
scientific user facility sponsored by the United States of America
Department of Energy under Contract DE-AC05-76RL0 1830. Development of modeling concepts was also supported by P41 EB023912 through H.M.S. at the Center for Reproducible
Biomedical Modeling.

\section{Author Contributions}\label{author-contributions}

H.S.W., S.E.B., K.E.B-J. and D.M. conceived and designed the study.
A.C., N.B.C. and S.J. added modeling and data analysis approaches, S.S.A
and H.M.S contributed ideas on modular frameworks and community
engagement and J.P.B and H.M.S contributed concepts on model validation
and benchmarking. H.S.W. wrote the first draft of the manuscript and
prepared the figure. All authors reviewed and edited the final
manuscript. No AI was used in the writing or editing of the paper, but
AI assisted in looking up references and preparing the figure.

\section{References}\label{references}

1. Brooks, S.M. and H.S. Alper, \emph{Applications, challenges, and
needs for employing synthetic biology beyond the lab.} Nature
Communications, 2021. \textbf{12}(1): p.~1390.

2. Brown, R.M., et al., \emph{Current nature-based biological practices
for rare earth elements extraction and recovery: Bioleaching and
biosorption.} Renewable and Sustainable Energy Reviews, 2023.
\textbf{173}: p.~113099.

3. Omer, R., et al., \emph{Engineered Bacteria-Based Living Materials
for Biotherapeutic Applications.} Front Bioeng Biotechnol, 2022.
\textbf{10}: p. 870675.

4. Nielsen, J. and J.D. Keasling, \emph{Engineering Cellular
Metabolism.} Cell, 2016. \textbf{164}(6): p.~1185-1197.

5. Dougherty, E.R. and I. Shmulevich, \emph{On the limitations of
biological knowledge.} Curr Genomics, 2012. \textbf{13}(7): p.~574-87.

6. Singh, R.S., \emph{Decoding `Unnecessary Complexity': A Law of
Complexity and a Concept of Hidden Variation Behind ``Missing
Heritability'' in Precision Medicine.} J Mol Evol, 2021. \textbf{89}(8):
p.~513-526.

7. Cheng, A., et al., \emph{Genetics Matters: Voyaging from the Past
into the Future of Humanity and Sustainability.} Int J Mol Sci, 2022.
\textbf{23}(7).

8. Esvelt, K.M. and H.H. Wang, \emph{Genome-scale engineering for
systems and synthetic biology.} Mol Syst Biol, 2013. \textbf{9}: p.~641.

9. Basler, G., \emph{Computational prediction of essential metabolic
genes using constraint-based approaches.} Methods Mol Biol, 2015.
\textbf{1279}: p. 183-204.

10. Sauro, H.M., et al., \emph{From FAIR to CURE: guidelines for
computational models of biological systems.} NPJ Syst Biol Appl, 2026.
\textbf{12}(1).

11. Meyer, P. and J. Saez-Rodriguez, \emph{Advances in systems biology
modeling: 10 years of crowdsourcing DREAM challenges.} Cell Syst, 2021.
\textbf{12}(6): p.~636-653.

12. Wiley, H.S., et al., \emph{A Roadmap for the Future of Systems
Biology in Cancer Research.} Cancer Res, 2025. \textbf{85}(24):
p.~4880-4889.

13. Bunne, C., et al., \emph{How to build the virtual cell with
artificial intelligence: Priorities and opportunities.} Cell, 2024.
\textbf{187}(25): p. 7045-7063.

14. Michalska, K. and A. Joachimiak, \emph{Structural genomics and the
Protein Data Bank.} Journal of Biological Chemistry, 2021. \textbf{296}:
p. 100747.

15. Stevens, R.C., S. Yokoyama, and I.A. Wilson, \emph{Global efforts in
structural genomics.} Science, 2001. \textbf{294}(5540): p.~89-92.

16. Burley, S.K., et al., \emph{Updated resources for exploring
experimentally-determined PDB structures and Computed Structure Models
at the RCSB Protein Data Bank.} Nucleic Acids Res, 2025.
\textbf{53}(D1): p. D564-D574.

17. Jumper, J., et al., \emph{Highly accurate protein structure
prediction with AlphaFold.} Nature, 2021. \textbf{596}(7873):
p.~583-589.

18. Dibaeinia, P., et al., \emph{Virtual Cells Need Context, Not Just
Scale.} bioRxiv, 2026.

19. Service, R.F., \emph{Structural biology. Protein structure
initiative: phase 3 or phase out.} Science, 2008. \textbf{319}(5870):
p.~1610-3.

20. Paulevé, L., et al., \emph{Reconciling qualitative, abstract, and
scalable modeling of biological networks.} Nature Communications, 2020.
\textbf{11}(1): p.~4256.

21. Zhu, K., et al., \emph{A unified approach to protein domain parsing
with inter-residue distance matrix.} Bioinformatics, 2023.
\textbf{39}(2).

22. Chothia, C. and A.M. Lesk, \emph{The relation between the divergence
of sequence and structure in proteins.} EMBO J, 1986. \textbf{5}(4):
p.~823-6.

23. Sousounis, K., et al., \emph{Conservation of the three-dimensional
structure in non-homologous or unrelated proteins.} Hum Genomics, 2012.
\textbf{6}(1): p.~10.

24. Caetano-Anolles, G., et al., \emph{Emergence of Hierarchical
Modularity in Evolving Networks Uncovered by Phylogenomic Analysis.}
Evol Bioinform Online, 2019. \textbf{15}: p.~1176934319872980.

25. Hatleberg, W.L. and V.F. Hinman, \emph{Modularity and hierarchy in
biological systems: Using gene regulatory networks to understand
evolutionary change.} Curr Top Dev Biol, 2021. \textbf{141}: p.~39-73.

26. Lorenz, D.M., A. Jeng, and M.W. Deem, \emph{The emergence of
modularity in biological systems.} Phys Life Rev, 2011. \textbf{8}(2):
p.~129-60.

27. Guo, M.G., D.N. Sosa, and R.B. Altman, \emph{Challenges and
opportunities in network-based solutions for biological questions.}
Brief Bioinform, 2022. \textbf{23}(1).

28. Alcala-Corona, S.A., et al., \emph{Modularity in Biological
Networks.} Front Genet, 2021. \textbf{12}: p.~701331.

29. Kaltenbach, H.M. and J. Stelling, \emph{Modular analysis of
biological networks.} Adv Exp Med Biol, 2012. \textbf{736}: p.~3-17.

30. Lill, D., et al., \emph{Mapping connections in signaling networks
with ambiguous modularity.} NPJ Syst Biol Appl, 2019. \textbf{5}: p.~19.

31. Goelzer, A. and V. Fromion, \emph{Resource allocation in living
organisms.} Biochem Soc Trans, 2017. \textbf{45}(4): p.~945-952.

32. Yizhak, K., et al., \emph{Integrating quantitative proteomics and
metabolomics with a genome-scale metabolic network model.}
Bioinformatics, 2010. \textbf{26}(12): p.~i255-60.

33. Lakrisenko, P. and D. Weindl, \emph{Dynamic models for metabolomics
data integration.} Current Opinion in Systems Biology, 2021.
\textbf{28}: p. 100358.

34. Saez-Rodriguez, J., et al., \emph{Automatic decomposition of kinetic
models of signaling networks minimizing the retroactivity among
modules.} Bioinformatics, 2008. \textbf{24}(16): p.~i213-9.

35. Prado-Velasco, M., \emph{Modular dynamics paradigm in biosystems
multilevel modeling: Software design and PBPK/PD validation.} Computers
in Biology and Medicine, 2025. \textbf{188}: p.~109856.

36. Montaño-Gutierrez, L.F., et al., \emph{Analysing and meta-analysing
time-series data of microbial growth and gene expression from plate
readers.} PLoS Comput Biol, 2022. \textbf{18}(5): p.~e1010138.

37. Krishnamurthi, V.R., et al., \emph{A new analysis method for
evaluating bacterial growth with microplate readers.} PLoS One, 2021.
\textbf{16}(1): p. e0245205.

38. Kim, M. and I. Tagkopoulos, \emph{Data integration and predictive
modeling methods for multi-omics datasets.} Molecular Omics, 2018.
\textbf{14}(1): p.~8-25.

39. Prinn, R.G., \emph{Development and application of earth system
models.} Proceedings of the National Academy of Sciences, 2013.
\textbf{110}(supplement\_1): p.~3673-3680.

40. Lee, J., et al., \emph{Systematic and objective evaluation of Earth
system models: PCMDI Metrics Package (PMP) version 3.} Geosci. Model
Dev., 2024. \textbf{17}(9): p.~3919-3948.

41. Banga, J.R. and A.F. Villaverde, \emph{Mechanistic dynamic modelling
of biological systems: The road ahead.} Current Opinion in Systems
Biology, 2025. \textbf{42}: p.~100553.

42. Sechkar, K., et al., \emph{A coarse-grained bacterial cell model for
resource-aware analysis and design of synthetic gene circuits.} Nature
Communications, 2024. \textbf{15}(1): p.~1981.

43. Andrews, S.S., H.S. Wiley, and H.M. Sauro, \emph{Design patterns of
biological cells.} Bioessays, 2024. \textbf{46}(3): p.~e2300188.

44. Mallavarapu, A., et al., \emph{Programming with models: modularity
and abstraction provide powerful capabilities for systems biology.} J R
Soc Interface, 2009. \textbf{6}(32): p.~257-70.

45. Li, X., et al., \emph{Modular deregulation of central carbon
metabolism for efficient xylose utilization in Saccharomyces
cerevisiae.} Nat Commun, 2025. \textbf{16}(1): p.~4551.

46. Pan, M., et al., \emph{Modular assembly of dynamic models in systems
biology.} PLoS Comput Biol, 2021. \textbf{17}(10): p.~e1009513.

47. Petersen, B.K., G.E. Ropella, and C.A. Hunt, \emph{Toward modular
biological models: defining analog modules based on referent
physiological mechanisms.} BMC Syst Biol, 2014. \textbf{8}: p.~95.

48. Del Vecchio, D., A.J. Dy, and Y. Qian, \emph{Control theory meets
synthetic biology.} Journal of The Royal Society Interface, 2016.
\textbf{13}(120).

49. Wu, S., et al., \emph{Recent Advances in Mass Spectrometry-Based
Protein Interactome Studies.} Molecular \& Cellular Proteomics, 2025.
\textbf{24}(1): p.~100887.

50. Chatterjee, A., et al., \emph{Improving the generalizability of
protein-ligand binding predictions with AI-Bind.} Nat Commun, 2023.
\textbf{14}(1): p.~1989.

51. Dörig, C., et al., \emph{Global profiling of protein complex
dynamics with an experimental library of protein interaction markers.}
Nature Biotechnology, 2025. \textbf{43}(9): p.~1562-1576.

52. Kadelka, C., et al., \emph{Modularity of biological systems: a link
between structure and function.} J R Soc Interface, 2023.
\textbf{20}(207): p.~20230505.

53. Klingenberg, C.P., \emph{Studying morphological integration and
modularity at multiple levels: concepts and analysis.} Philos Trans R
Soc Lond B Biol Sci, 2014. \textbf{369}(1649): p.~20130249.

54. Masison, J., et al., \emph{A modular computational framework for
medical digital twins.} Proc Natl Acad Sci U S A, 2021.
\textbf{118}(20).

55. Agmon, E., et al., \emph{Vivarium: an interface and engine for
integrative multiscale modeling in computational biology.}
Bioinformatics, 2022. \textbf{38}(7): p.~1972-1979.

56. Schaffer, L.V. and T. Ideker, \emph{Mapping the multiscale structure
of biological systems.} Cell Systems, 2021. \textbf{12}(6): p.~622-635.

57. van Aalst, M., et al., \emph{MxlPy---Python package for mechanistic
learning and hybrid modelling in life science.} Bioinformatics Advances,
2025. \textbf{5}(1): p.~vbaf294.

58. Johnson, J.A.I., et al., \emph{Human interpretable grammar encodes
multicellular systems biology models to democratize virtual cell
laboratories.} Cell, 2025. \textbf{188}(17): p.~4711-4733.e37.

59. Lewis, N.E., H. Nagarajan, and B.O. Palsson, \emph{Constraining the
metabolic genotype--phenotype relationship using a phylogeny of in
silico methods.} Nature Reviews Microbiology, 2012. \textbf{10}(4):
p.~291-305.

60. Passi, A., et al., \emph{Genome-Scale Metabolic Modeling Enables
In-Depth Understanding of Big Data.} Metabolites, 2021. \textbf{12}(1).

61. Bordbar, A., et al., \emph{Constraint-based models predict metabolic
and associated cellular functions.} Nat Rev Genet, 2014. \textbf{15}(2):
p. 107-20.

62. Kerkhoven, E.J., \emph{Advances in constraint-based models: methods
for improved predictive power based on resource allocation constraints.}
Current Opinion in Microbiology, 2022. \textbf{68}: p.~102168.

63. Chen, Y. and J. Nielsen, \emph{Energy metabolism controls phenotypes
by protein efficiency and allocation.} Proc Natl Acad Sci U S A, 2019.
\textbf{116}(35): p.~17592-17597.

64. Li, F., et al., \emph{Deep learning-based kcat prediction enables
improved enzyme-constrained model reconstruction.} Nature Catalysis,
2022. \textbf{5}(8): p.~662-672.

65. Philipps, M., N. Schmid, and J. Hasenauer, \emph{Current state and
open problems in universal differential equations for systems biology.}
NPJ Syst Biol Appl, 2025. \textbf{11}(1): p.~101.

66. Joslin, E.J., et al., \emph{Structure of the EGF receptor
transactivation circuit integrates multiple signals with cell context.}
Mol Biosyst, 2010. \textbf{6}(7): p.~1293-306.

67. Taylor, R.C., et al., \emph{Changes in translational efficiency is a
dominant regulatory mechanism in the environmental response of
bacteria.} Integrative Biology, 2013. \textbf{5}(11): p.~1393-1406.

68. Ma, J., et al., \emph{Using deep learning to model the hierarchical
structure and function of a cell.} Nat Methods, 2018. \textbf{15}(4): p.
290-298.

69. Hucka, M., et al., \emph{Promoting Coordinated Development of
Community-Based Information Standards for Modeling in Biology: The
COMBINE Initiative.} Front Bioeng Biotechnol, 2015. \textbf{3}: p.~19.

70. Singla, J. and K.L. White, \emph{A community approach to whole-cell
modeling.} Current Opinion in Systems Biology, 2021. \textbf{26}:
p.~33-38.

71. Keating, S.M., et al., \emph{SBML Level 3: an extensible format for
the exchange and reuse of biological models.} Mol Syst Biol, 2020.
\textbf{16}(8): p.~e9110.

72. Mogilner, A., R. Wollman, and W.F. Marshall, \emph{Quantitative
Modeling in Cell Biology: What Is It Good for?} Developmental Cell,
2006. \textbf{11}(3): p.~279-287.

73. Tummler, K., et al., \emph{New types of experimental data shape the
use of enzyme kinetics for dynamic network modeling.} FEBS J, 2014.
\textbf{281}(2): p.~549-71.

74. Sahin, A., D.R. Weilandt, and V. Hatzimanikatis, \emph{Optimal
enzyme utilization suggests that concentrations and thermodynamics
determine binding mechanisms and enzyme saturations.} Nat Commun, 2023.
\textbf{14}(1): p.~2618.

75. Leek, J.T., et al., \emph{Tackling the widespread and critical
impact of batch effects in high-throughput data.} Nat Rev Genet, 2010.
\textbf{11}(10): p.~733-9.

76. Park, J.M., T.Y. Kim, and S.Y. Lee, \emph{Prediction of metabolic
fluxes by incorporating genomic context and flux-converging pattern
analyses.} Proc Natl Acad Sci U S A, 2010. \textbf{107}(33): p.~14931-6.

77. Ferreira, M.A.M., W.B.D. Silveira, and Z. Nikoloski, \emph{Protein
constraints in genome-scale metabolic models: Data integration,
parameter estimation, and prediction of metabolic phenotypes.}
Biotechnol Bioeng, 2024. \textbf{121}(3): p.~915-930.

78. Sielemann, K., A. Hafner, and B. Pucker, \emph{The reuse of public
datasets in the life sciences: potential risks and rewards.} PeerJ,
2020. \textbf{8}: p.~e9954.

79. Gruber, C.H., et al., \emph{Systematic identification of allosteric
effectors in Escherichia coli metabolism.} Proc Natl Acad Sci U S A,
2025. \textbf{122}(10): p.~e2423767122.

80. Peng, H., et al., \emph{Ligand interaction landscape of
transcription factors and essential enzymes in E. coli.} Cell, 2025.
\textbf{188}(5): p. 1441-1455 e15.

81. Zhang, J., et al., \emph{Combining mechanistic and machine learning
models for predictive engineering and optimization of tryptophan
metabolism.} Nature Communications, 2020. \textbf{11}(1): p.~4880.

82. Sarker, I.H., \emph{AI-Based Modeling: Techniques, Applications and
Research Issues Towards Automation, Intelligent and Smart Systems.} SN
Comput Sci, 2022. \textbf{3}(2): p.~158.

83. Alseekh, S., et al., \emph{Mass spectrometry-based metabolomics: a
guide for annotation, quantification and best reporting practices.} Nat
Methods, 2021. \textbf{18}(7): p.~747-756.

84. Buergel, T., et al., \emph{Metabolomic profiles predict individual
multidisease outcomes.} Nat Med, 2022. \textbf{28}(11): p.~2309-2320.

85. Meier, F., et al., \emph{diaPASEF: parallel accumulation-serial
fragmentation combined with data-independent acquisition.} Nat Methods,
2020. \textbf{17}(12): p.~1229-1236.

86. Mann, M., et al., \emph{The Coming Age of Complete, Accurate, and
Ubiquitous Proteomes.} Molecular Cell, 2013. \textbf{49}(4): p.~583-590.

87. Wenk, D., et al., \emph{Recent developments in
mass-spectrometry-based targeted proteomics of clinical cancer
biomarkers.} Clin Proteomics, 2024. \textbf{21}(1): p.~6.

88. Gao, S., et al., \emph{Empowering biomedical discovery with AI
agents.} Cell, 2024. \textbf{187}(22): p.~6125-6151.

89. Driscoll, M.K. and A. Zaritsky, \emph{Data science in cell imaging.}
J Cell Sci, 2021. \textbf{134}(7).

90. Hajnajafi, K. and M.A.~Iqbal, \emph{Mass-spectrometry based
metabolomics: an overview of workflows, strategies, data analysis and
applications.} Proteome Sci, 2025. \textbf{23}(1): p.~5.

91. Packalen, M. and J. Bhattacharya, \emph{NIH funding and the pursuit
of edge science.} Proceedings of the National Academy of Sciences, 2020.
\textbf{117}(22): p.~12011-12016.

92. von Meijenfeldt, F.A.B., P. Hogeweg, and B.E. Dutilh, \emph{A social
niche breadth score reveals niche range strategies of generalists and
specialists.} Nature Ecology \& Evolution, 2023. \textbf{7}(5):
p.~768-781.

93. McClure, R., et al., \emph{Interaction Networks Are Driven by
Community-Responsive Phenotypes in a Chitin-Degrading Consortium of Soil
Microbes.} mSystems, 2022. \textbf{7}(5): p.~e00372-22.

94. Mahilkar, A., et al., \emph{Experimental Evolution of Anticipatory
Regulation in Escherichia coli.} Front Microbiol, 2021. \textbf{12}: p.
796228.

95. Mao, J., et al., \emph{Relieving metabolic burden to improve
robustness and bioproduction by industrial microorganisms.}
Biotechnology Advances, 2024. \textbf{74}: p.~108401.

96. Gottschalk, U., K. Brorson, and A.A. Shukla, \emph{The need for
innovation in biomanufacturing.} Nat Biotechnol, 2012. \textbf{30}(6):
p. 489-92.

97. Mitchell, A., et al., \emph{Adaptive prediction of environmental
changes by microorganisms.} Nature, 2009. \textbf{460}(7252):
p.~220-224.

98. Olsson, L., et al., \emph{Robustness: linking strain design to
viable bioprocesses.} Trends in Biotechnology, 2022. \textbf{40}(8):
p.~918-931.

99. Moult, J., et al., \emph{A large-scale experiment to assess protein
structure prediction methods.} Proteins, 1995. \textbf{23}(3): p.~ii-v.

100. Kryshtafovych, A., et al., \emph{Critical assessment of methods of
protein structure prediction (CASP)---Round XV.} Proteins: Structure,
Function, and Bioinformatics, 2023. \textbf{91}(12): p.~1539-1549.

101. Eyring, V., et al., \emph{Overview of the Coupled Model
Intercomparison Project Phase 6 (CMIP6) experimental design and
organization.} Geosci. Model Dev., 2016. \textbf{9}(5): p.~1937-1958.

102. Launius, R.D., \emph{The National Labs: Science in an American
System, 1947--1974. By Peter J. Westwick. (Cambridge: Harvard University
Press, 2003. xii, 403 pp.~\$49.95, isbn 0-674-00948-7.).} Journal of
American History, 2004. \textbf{91}(1): p.~303-304.

\end{document}